\documentstyle [12pt] {article}

\catcode`@=12
\begin{document}
\vspace{0.5in}
\oddsidemargin -.375in
\newcount\sectionnumber
\sectionnumber=0
\def\be{\begin{equation}}
\def\ee{\end{equation}}
\begin{flushright} UH-511-787-94\\March 1994\\hep-ph/9403360\
\end{flushright}
\vspace {.5in}
\begin{center}
{\Large\bf Gravitational uncertainties from dimension six \\ operators
on supersymmetric GUT predictions\\}
\vspace{.5in}
{\bf Alakabha Datta${}^{a)}$},
{\bf Sandip Pakvasa${}^{a)}$} and {\bf Utpal Sarkar${}^{b)}$\\}
\vspace{.1in}
${}^{a)}$ {\it
Physics Department, University of Hawaii at Manoa, 2505 Correa
Road, Honolulu, HI 96822, USA.}\\
${}^{b)}$ {\it Theory Group,
Physical Research Laboratory, Ahmedabad - 380009, India.}\\
\vskip .5in
{\bf Abstract}
\end{center}

\vskip .1in
\begin{quotation}

We consider the gravity induced dimension six terms in addition to
 the dimension
five terms in the SUSY GUT Lagrangian
 and find that the prediction for $\alpha_s$ may be washed
out completely in supersymmetric grand unified theories unless
the triplet higgs mass is smaller than
$ 7\times 10^{16} $ GeV.

\end{quotation}
\vskip .5in

Recently, Hall and Sarid,$\mbox{}^1$ and Langacker and
Polonsky$\mbox{}^2$ have shown that the prediction of the
strong coupling constant $\alpha_s$ in the minimal
supersymmetric $SU(5)$ grand unified theory is smeared out when
dimension five non-renormalizable operators arising from
gravity is included ( Recently Planck scale effects have also been considered
by A.Vayonakis$\mbox{}^2$).
 In this brief report we point out that for high
GUT scale higher dimensional operators can be as significant  as
dimension five operators. In particular
we show that these operators
 can wash out the prediction
for $\alpha_s$ completely.

In the case of non-supersymmetric GUTs it was shown$\mbox{}^3$ that by
considering dimension five operators alone it is not possible
to make minimal $SU(5)$ GUT consistent with the LEP data and
proton decay limit. Whereas by considering both dimension five
and dimension six operators one can make the minimal $SU(5)$
GUT consistent with LEP data and satisfy the proton decay limit.$\mbox{}^4$

We use the notation of Hall and Sarid and include the GUT
threshold corrections to compare our result with that of Ref.1.
 We include both dimension 5 and dimension 6 operators,
which might originate from non-renormalizable quantum gravity
effect, and write
\begin{equation}
\delta {\cal L} = \displaystyle \frac{c}{2 \hat{M}_P} {\rm tr}
( G G \Sigma) + \displaystyle \frac{1}{2 \hat{M}_P^{2}}[ d_{11}{ \frac{1}{2}
{\rm tr} (G \Sigma^2 G) +d_{12} \frac{1}{2} {\rm tr} (G \Sigma G \Sigma)} +
d_{2} {\rm tr} (G^2) {\rm tr} (\Sigma^2) + d_{3}  {\rm tr} (G \Sigma)
{\rm tr} (G \Sigma) ]\
\end{equation}
where ${ \hat{M}_P}=(8\pi G_N)^{-1/2}\simeq 2.4\times 10^{18}$ GeV
 is the reduced Planck mass.

 Then these terms will modify the kinetic
energy terms of the standard model gauge bosons to
\begin{eqnarray}
{\cal L}_{gauge} &=& - \frac{1}{4} (F F)_{U(1)} \left[ 1 +
\frac{c}{2} \frac{v}{\hat{M}_P} \left( - \frac{1}{2 \sqrt{15}}
\right) +  \frac{v^2}{2 \hat{M}_P^2} \left(
\frac{1}{15} \right) \left( d_{1} \frac{7}{4} +d_{2} \frac{15}{2}
+ d_{3} \frac{15}{2}
\right) \right]\nonumber\\
 & &- \frac{1}{4} (F F)_{SU(2)} \left[ 1 +
\frac{c}{2} \frac{v}{\hat{M}_P} \left( - \frac{3}{2 \sqrt{15}}
\right) +  \frac{v^2}{2 \hat{M}_P^2} \left(
\frac{1}{15} \right) \left(d_{1} \frac{9}{4} + d_{2} \frac{15}{2}
\right) \right]\nonumber\\
 & &- \frac{1}{4} (F F)_{SU(3)} \left[ 1 +
\frac{c}{2} \frac{v}{\hat{M}_P} \left(  \frac{1}{\sqrt{15}}
\right) +  \frac{v^2}{2 \hat{M}_P^2} \left(
\frac{1}{15} \right) \left(d_{1} +d_{2} \frac{15}{2}\right) \right]
\end{eqnarray}
where we have defined $d_{1}=(d_{11}+d_{12})/{2}$ as the the first two
operators in eqn.(1) always contribute equally.
Note that in principle one can also include operators of dimensions higher
than six in our analysis but  their contributions to ${\vec{\epsilon}}$, where
  ${\vec{\epsilon}} {\alpha_{G}}^{-1}$ is the amount by which
 $ {\vec{\alpha_{G}}}^{-1}$ gets modified in the evolution equations for the
coupling constants,
can be included by absorbing them in the co-efficients $d_{1},d_{2}$ and
$d_{3}$. Since we are interested only in gauge coupling evolutions it is
 thus sufficient to confine our analysis to just dimension five and dimension
 six operators
for minimal supersymmetric $SU(5)$ GUT and see how they
can affect the predictions of $\alpha_s$.
 At the one loop level the gauge coupling, evaluated at the $Z$ mass
$\vec{\alpha}^{-1} \equiv \vec{\alpha}^{-1}(m_Z) \equiv
( {{\alpha}_1}^{-1}, {{\alpha}_2}^{-1},
{{\alpha}_3}^{-1})$ will be related to the GUT scale $(M_G)$
gauge coupling constant
$$ {\vec{\alpha}}^{-1} = {\alpha}_G^{-1} \left( \vec{1} +
\vec{\epsilon}_5 + \vec{\epsilon}_6 \right) - \sum_a
\vec{\beta}_a \ln \left( M_a \over M_G \right) $$ where,
$\vec{\epsilon}_5 \equiv \frac{c}{2} \frac{v}{\hat{M}_P}
\left(  \frac{1}{\sqrt{15}}  \right) \left( - \frac{1}{2}, -
\frac{3}{2}, 1 \right)$ and $\vec{\epsilon}_6 \equiv
\frac{d}{30} \frac{v^2}{\hat{M}_P^2} \left(d_{1} \frac{7}{4} +
d_{2} \frac{15}{2}
+ d_{3} \frac{15}{2}, d_{1} \frac{9}{4} + d_{2} \frac{15}{2},
d_{1} +d_{2} \frac{15}{2}
 \right).$

Then, following Hall and Sarid$\mbox{}^1$ the
modified unification equations are given by
\begin{eqnarray}
\frac{2}{\alpha_s} + \frac{6}{5 \pi} \ln \frac{M_{tr}}{m_Z} -
\sqrt{\frac{12}{5}} \frac{c}{2} \frac{v}{\hat{M}_P}
\frac{1}{{\alpha}_G} + [\frac{1}{5} d_{1} -\frac{1}{2} d_{3}]
\frac{v^2}{2\hat{M}_P^2} \frac{1}{{\alpha}_G} & = & f_1(s^2,
m_0, m_{1 \over 2}, \mu, m_H) \nonumber \\
\frac{2}{\alpha_s} + \frac{9}{\pi} \ln \frac{5}{12} +
\frac{12}{\pi} \ln g_5 + \frac{6}{\pi} \ln \lambda_{24} +
\frac{18}{\pi} \ln \frac{v}{m_Z} + \frac{5}{2} \frac{d_{3}}{2}
\frac{v^2}{\hat{M}_P^2} \frac{1}{{\alpha}_G} &=& f_2(s^2, m_0,
m_{1 \over 2})
 \end{eqnarray}
 where $M_{tr}$ is the mass of the color triplet higgs.

Subtracting one of the equations in (3) from the other we obtain an equation
for $M_{tr}$ which can be written as
\begin{eqnarray}
\frac{-84}{5{\pi}}\ln t &= & w_1t^2+w_2t+b\
 \end{eqnarray}
where
\begin{eqnarray*}
 t={\frac{M_{tr}}{\hat{M}_P}} \ , \ w_1=\frac{1}{\alpha_{G}{{\lambda}_5}^2}
 [\frac{18}{5} d_{3} - \frac{6}{25} d_{1}] \ , \
  w_2=\frac{6}{5 \alpha_{G}{\lambda}_5}c \\
  b=f_1-f_2 +
 \frac{6}{\pi} \ln \frac{\lambda_{24}}{{\lambda_{5}}^3}+ \frac{6}{\pi}
\ln4{\pi}{\alpha_{G}} + \frac{84}{5{\pi}}\ln\frac{\hat{M}_P}{m_Z}
\end{eqnarray*}

Defining, $x = {M_{tr}}/{\lambda_5\hat{M}_P} $ we can rewrite
the first eqn in (3) as
\begin{eqnarray}
\frac{2}{\alpha_s}&=& f_1(s^2,m_0,m_{1\over2},\mu,m_H)-
\frac{6}{5\pi}\ln\frac{M_{tr}}{m_Z}+\frac{6}{5}\frac{c}{\alpha_G}x
+[\frac{3}{5} d_{3} - \frac{6}{25} d_{1}]\frac{x^2}{\alpha_G} \
\end{eqnarray}

We now numerically solve eqn.(4) for $t$ and then use eqn.(5) to calculate
$\alpha_{s}$. We use the same mass spectrum  and  ranges of
parameters ($s^2$,
$m_0$, $m_{1 \over 2}$, $\mu$, $m_{{H}_{2}}$,$\lambda_{5}$,$\lambda_{24}$,$c$)
as in Ref.1. In other words we vary the light superpartner masses and the
second higgs doublet mass between 100 GeV and 1 TeV, $s^2$ between 0.2314 and
0.2324 $ \mbox{}^5$, $\lambda_{5}$ and $\lambda_{24}$ between $.1$ and $3$
while we
constrain $|c|< 1$. The co-effecients $d_{1}$, $d_{2}$ and $d_{3}$ are
unknown,
but we see from eqn.(4) and eqn.(5) that only $d_{1}$ and $d_{3}$ contribute
 to the equations for $M_{tr}$ and $\alpha_{s}$. We also observe from eqn.(4)
that $d_{3}$ has a much larger coefficient
. We can now consider two scenarios, one with
$|d_{1}|< 1;d_{3}=0$ and
$|d_{1}|= 0; |d_{3}|< 1$.
There may be multiple solutions to eqn.(3) and we have chosen
the lowest solution in our analysis. To select the lowest solution we define
two critical solutions $t_{1}$ and $t_{2}$ which are given by
\begin{eqnarray}
t_{1} &=& \frac{t_{ex}}{2}[1+\sqrt{(1-2y)}] \\
t_{2} &=& \frac{t_{ex}}{2}[1-\sqrt{(1-2y)}]
\end{eqnarray}
where $t_{ex}={-w{_2}}/{2w{_1}}$ ,  $ a=84/5\pi$ and
 $y={a}/{w{_1}t_{ex}^2}$.
For $w_{2}=0$ we have one critical solution $t_{cr}$ given by

\begin{eqnarray}
t_{cr} & =& \sqrt{\frac{-a}{2w_{1}}}
\end{eqnarray}
The critical solutions correspond to points where the tangent to the
logarathmic
function on the left hand side of eqn.(4) equals the tangent to the parabola
on the right hand side of eqn.(4).
When $t_{1}$ and $t_{2}$ are both real and positive and distinct
 from one another  we can have
 at most three
solutions, one below $t_{1}$, one between $t_{1}$
 and $t_{2}$ and one above $t_{2}$. If instead the critical solutions are
 real and positive,  but equal then we can have
at most two solutions. For $w_{1} < 0$ there
 is always one real positive
critical solution and so there can be up to two solutions one on either side
of the critical solution.
When there is no real, positive critical solution there can be up to one
solution to eqn.(4).
For $w_{1}=0$, as observed in Ref.1, there can be only
one solution for $w_{2}$ greater than $0$ while for $w_{2}$ less than $0$ there
can be upto two solutions lying on either side of the critical solution
$t_{critical}={-a}/{w_{2}}$.

{\bf Results}
For the case where
$|d_{1}|< 1;d_{3}=0$, the effect of dimension 6 operators are found to be
negligible. However for the case where
$|d_{1}|= 0; |d_{3}|< 1$, the effect of dimension six operator
 can be significant.
 In fig.1 (a) we show a plot of the solutions in the
$\alpha_{s}- M_{tr}$ plane. Although we cut off the figure at
$\alpha_{s} = 1 $, we mention that there are solutions for larger values
 of
$\alpha_{s}$ {\footnote{Of course, the
equations themselves cease to be valid if $\alpha_s$ is too
large.}}. In table.1 we show the ranges of
$\alpha_{s}$ for different $M_{tr}$. Fig.1 (b) is a blow up of fig.1 (a) for
$\alpha_s \le 0.12$. Here, we have used
 a much smaller grid size for $\lambda_{5}$ in our numerical computation; as
a result,
 some solutions that do not show up in fig.1 (a) now appear in fig.1 (b).
 We observe that for $M_{tr} \ge 7\times10^{16}$
GeV the range
of the solutions for $\alpha_{s}$ is greatly increased. We also note that with
dimension 6 operators it is now possible to get values of $\alpha_{s}$ below
0.11 which was not possible with pure dimension 5 term. This could be of
interest if  in the future the central value of
 $\alpha_s = 0.120 \pm 0.007 \pm
0.002 $ $\mbox{}^5$ shifts down by $ \sim 1.5 \sigma $. ( It is interesting to
note that
 such
a low value of $\alpha_s$ $ (0.108 \pm 0.004) $ is indeed obtained in an
 analysis
 of
LEP data by Maxwell et al. $\mbox{}^6$ where it is claimed that the standard
perturbative QCD analyses used to extract $\alpha_s$ from LEP data
do not correctly take into account higher order NNLO corrections which can
be sizeable for some of the LEP observables used in the determination of
$\alpha_s$.)
  We found that
 solutions with large values of $\alpha_{s}$ and small values
 of $\alpha_s$ ( less than 0.11) correspond
to small values of $\lambda_{5}$ in the range 0.1 to 0.3 indicating a high
value for $M_{X}$ (or $x$) and consequently large gravitational corrections.
 When the unification scale is close to the Planck
scale the
magnitude of the terms induced by the
higher dimensional operators in eqn.(5)
can become comparable to the combination
of the first two terms,  resulting in a much wider range
for $ \alpha_s $.
 In our calculations we have constrained
  the heavy masses to be less than
 $\hat{M}_P $.
 To compare to the results with
only the dimension five operator included, we note
that in that case,  the parameter
 $x$ always is of the order of $10^{-2}$. However
 the inclusion of the dimension
six operators allows $ x $  to be an order of magnitude
 higher indicating a higher unification scale close to $ \hat{M}_P $
(Note $\frac{M_X}{\hat{M}_P}=\sqrt{8\pi\alpha_G}x \sim x $ for $\alpha_G=
\frac{1}{25} $; where $ M_X $ is the vector boson mass) and therefore it
 is not
 surprising that the effects of the
 higher dimensional operators are significant.

In summary, we have shown that the inclusion of dimension 6 operators
may totally wash out the predictions for the strong coupling constant
and further,  that the correlation between  $\alpha_s $ and $ M_{tr} $
is also destroyed unless we constrain the triplet higgs mass $M_{tr}
< 7\times 10^{16} $ GeV because
 as we see from Table.1 the range of $\alpha_s$
increases significantly from the point
$M_{tr}
= 7\times 10^{16}$ GeV onwards. Turning this around, if we require that
SUSY-GUT make calculable predictions at the electroweak scale in
the presence of gravity induced non-renormalizable operators we may
infer more restrictive bounds on the triplet higgs mass than are
available in the literature $\mbox{}^7$.
\begin{table}
\caption{Allowed ranges of $\alpha_s$ for various $M_{tr}$ for $|d_1|=0 \ ;\
|d_3|<1$}
\begin{center}
\begin{tabular}{|c|c|c|}
\hline
$M_{tr}\times 10^{16}$ GeV  & $\alpha_s$(max) & $\alpha_s$(min)  \\
\hline
$1$  & 0.144 & 0.115  \\
\hline
$2$  & 0.146 & 0.118  \\
\hline
$3$  & 0.146 & 0.119  \\
\hline
$4$  & 0.147 & 0.120  \\
\hline
$5$  & 0.147 & 0.119  \\
\hline
$6$  & 0.147 & 0.120  \\
\hline
$7$  & 0.404 & 0.124  \\
\hline
$8$  & 0.705 & 0.121  \\
\hline
$9$  & 1.408 & 0.121  \\
\hline
$10$  & 2.95 & 0.121  \\
\hline
$14$  & 2.82& 0.101  \\
\hline
$18$  & 1.742 & 0.0660 \\
\hline
$22$  & 3.89 & 0.0590  \\
\hline
$26$  & 3.82 & 0.0570  \\
\hline
$30$  & 3.36 & 0.0560 \\
\hline
\end{tabular}
\end{center}
\end{table}

{\bf Acknowledgement}
 We would like to thank Prof.Xerxes
Tata for useful discussions and Dr.Atanu Basu for computer assistance.
This work was supported in part  by US
D.O.E under contract DE-AMO3-76SF-00325

\subsection{ Figure Captions}
\begin{itemize}
\item {\bf fig.1 (a)}: The predictions for $\alpha_s$ in minmal SU(5) SUSY GUT
as a function of the color-triplet higgs mass $ M_{tr}$ in GeV.
\item {\bf fig.1 (b)}: Predictions for $\alpha_s$ below 0.12. The numerical
calculations for this figure is done with a smaller grid size for
$\lambda_5$ than was used for fig.1 (a).
 \end{itemize}
\end{document}